\documentclass[journal,twoside]{IEEEtranTCOM}

\usepackage{blindtext}

\usepackage[latin1]{inputenc}

\usepackage[noadjust]{cite}
\usepackage{graphicx}
\usepackage{psfrag}
\usepackage{subfig}
\usepackage{amsmath,amssymb,amsthm,mathrsfs,amsfonts,dsfont}
\usepackage{color}
\usepackage[algoruled]{algorithm2e}

\usepackage{fixltx2e}

\usepackage{multirow}
\usepackage{multicol}
\usepackage[hidelinks]{hyperref}

\newcommand{\diag}{\mathop{\mathrm{{\bf diag}}}}

\usepackage{amsthm}
\newtheoremstyle{note}{3pt}{3pt}{\itshape}{}{\itshape}{:}{.5em}{}
\theoremstyle{note}

\usepackage{acronym}
\newacro{PLC}{power line communication}
\newacro{OFDM}{orthogonal frequency-division multiplexing}
\newacro{HS-OFDM}{Hermitian symmetric OFDM}
\newacro{PSD}{power spectral density}
\newacro{BER}{bit error rate}
\newacro{SER}{symbol error rate}
\newacroindefinite{SER}{an}{a}
\newacro{SCRA}{spectral compressive resource allocation}
\newacro{TCRA}{temporal compressive resource allocation}
\newacro{STCRA}{spectral-temporal compressive resource allocation}
\newacro{nSNR}{normalized signal-to-noise ratio}
\newacroindefinite{nSNR}{an}{a}
\newacro{CFR}{channel frequency response}
\newacro{LTV}{linear time-variant}
\newacro{LPTV}{linear periodically time variant}
\newacroindefinite{LPTV}{an}{a}
\newacro{LTI}{linear time-invariant}
\newacroindefinite{LTI}{an}{a}
\newacro{DFT}{discrete Fourier transform}
\newacro{AWGN}{additive white Gaussian noise}
\newacro{QAM}{quadrature amplitude modulation}
\newacro{SNR}{signal-to-noise ratio}
\newacroindefinite{SNR}{an}{a}
\newacro{FT}{Fourier transform}
\newacro{CCDF}{complementary cumulative distribution function}
\newacro{CIR}{channel impulse response}
\newacro{AIC}{Akaike information criterion}
\newacro{BIC}{Bayesian information criterion}
\newacro{EDC}{efficient determination criterion}
\newacro{PDF}{probability density function}
\newacro{GMD}{Gaussian mixture distribution \acused{GMDs}}
\newacro{GMDs}{Gaussian mixture distributions \acused{GMD}}
\newacro{GC}{Gaussian component}
\newacro{CP}{component proportion}
\newacro{CSI}{channel state information}

\newacro{FDM}{frequency-division multiplexing}
\newacro{NBI}{narrowband interference}
\newacro{DSL}{digital subscriber line}
\newacro{RA}{rate-adaptive}
\newacro{MA}{margin-adaptive}
\newacro{DMT}{discrete multitone modulation}
\newacro{ZF}{zero-forcing}
\newacro{PAM}{pulse amplitude modulation}
\newacro{WF}{water filling}
\newacro{WSS}{wide-sense stationary}
\newacro{ISI}{intersymbol interference}
\newacro{QoS}{quality of service}

\newacro{TL}{transmission line}
\newacro{DL}{distribution line}
\newacro{HV}{high voltage}
\newacro{MV}{medium voltage}
\newacro{LV}{low voltage}
\newacro{TEM}{transverse electromagnetic}
\newacro{SG}{smart grid}
\newacro{IDFT}{inverse discrete Fourier transform}
\newacro{IFT}{inverse Fourier transform}
\newacro{HIF}{high impedance fault}
\newacro{LIF}{low impedance fault}
\newacro{TDR}{time-domain reflectometry}
\newacro{FDR}{frequency-domain reflectometry}

\newacro{IDTFT}{inverse discrete-time Fourier transform}
\newacro{PLM}{power line modem}
\newacro{TDMA}{time-division multiple access}
\newacro{FDMA}{frequency-division multiple access}
\newacro{DFnT}{discrete Fresnel transform}
\newacro{IDFnT}{inverse discrete Fresnel transform}
\newacro{OCDM}{orthogonal chirp-division multiplexing}

\usepackage{dblfloatfix}

\usepackage{accents}

\newacro{TDMA}{time-division multiple access}
\newacro{FDMA}{frequency-division multiple access}
\newacro{CDMA}{code-division multiple access}
\newacro{PSLR}{peak-to-sidelobe level ratio}
\newacro{ISLR}{integrated-sidelobe level ratio}

\newacro{SINR}{signal-to-interference-plus-noise ratio}
\newacro{NB}{narrowband}
\newacro{BB}{broadband}

\newacro{BPSK}{binary phase-shift keying}
\newacro{MTL}{multiconductor transmission line}

\newacro{MCTDR}{multicarrier time-domain reflectometry}
\newacro{OMTDR}{orthogonal multitone time-domain reflectometry}
\newacro{EMC}{electromagnetic compatibility}

\usepackage{multirow}
\usepackage[table,xcdraw]{xcolor}

\usepackage{flushend}

\begin{document}
\title{Orthogonal Chirp Division Multiplexing
	for Power Line Sensing via Time-Domain Reflectometry}

\author{Lucas Giroto de Oliveira,~\IEEEmembership{Student Member,~IEEE},
	Mateus de L. Filomeno,\\
	and Mois\'es V. Ribeiro,~\IEEEmembership{Senior Member,~IEEE}
	\thanks{Manuscript received month day, year; revised month day, year. This study was financed in part by the Coordena\c{c}\~ao de Aperfei\c{c}oamento de Pessoal de N\'ivel Superior - Brasil (CAPES) - Finance Code 001, Copel Distribui\c{c}\~ao LTD - PD 2866-0420/2015 , CNPq, FAPEMIG, INERGE and Smarti9 LTD.}
	\thanks{Lucas Giroto de Oliveira and Mateus de L. Filomeno are with the Electrical Engineering Department, Federal University of Juiz de Fora (UFJF), Juiz de Fora, Brazil  (e-mail: lgiroto@ieee.org, mateus.lima@engenharia.ufjf.br).}
	\thanks{Mois\'es V. Ribeiro is with the Electrical Engineering Department, Federal University of Juiz de Fora (UFJF), Juiz de Fora, Brazil, and Smarti9 Ltd, Brazil  (e-mail: mribeiro@ieee.org).}}


\maketitle

\begin{abstract}
	In this study, a time-domain reflectometry (TDR) system based on a baseband version of the orthogonal chirp-division multiplexing (OCDM) scheme, which is relies on a modified discrete Fresnel transform (DFnT), is proposed for power line sensing. After a detailed description of the system model, a multiple access scheme that exploits the contolution theorem of the modified DFnT for enabling distributed reflectometric and transferometric sensing of the monitored power line. Considering typical European underground low-voltage and US overhead medium-voltage (MV) power distribution networks, range resolution and maximum unambiguous range for the measurements are assessed for the proposed scheme. Next, a comparison with multiple access schemes based on the Hermitian symmetric orthogonal frequency-division multiplexing (HS-OFDM) discussed in a previous work is performed considering a Brazilian MV overhead scenario, being the number of measurements obtained over time, as well as signal-to-interference-plus-noise ratio (SINR) used as performance metrics. The attained results show that the proposed multiple access scheme results in the same range resolution as the others. Also, the highest number of measurements over time is obtained by the proposed scheme, which produces orthogonality among signals transmitted by different power line modems (PLMs) neither and time nor frequency domains, but rather in the Fresnel domain. Meanwhile, although its yielded SINR values are fair among the PLMs consisting the distributed sensing system, the proposed scheme is slightly outperformed by the HS-OFDM based on code-division multiple access and by the HS-OFDM based on frequency-division multiple access at some PLMs.
\end{abstract}


\IEEEpeerreviewmaketitle

\section{Introduction}\label{sec:introduction}

In the context of modern power line diagnosis, the use of higher frequencies than the mains has been widely considered in recently proposed techniques. A main reason for that is the higher amount of information that can be extracted from the monitored network by monitoring harmonic voltages and currents \cite{sedighizadeh2010,ghaderi2017}, or even traveling waves \cite{auzanneau2013} in a wider frequency band, allowing more efficient sensing of anomalies such as \ac{HIF} and cable degradation.

For enabling travelling wave-based approaches, the use of \ac{PLC} technology is an interesting alternative that has been recently considered \cite{milioudis2015}. Topics ranging from adjustments to \acp{PLM} for enabling network sensing \cite{passerini2018_2} to machine learning-aided power line monitoring \cite{huo2018} have been investigated. Also, reflectometric \cite{parte1} and transferometric \cite{auzanneau2016} sensing of power lines have been given attention.

In this context, \ac{TDR} techniques such as \ac{MCTDR} \cite{naik2006,amini2009,lelong2009} and its \ac{OFDM}-based version \ac{OMTDR} \cite{hassen2015} allow effective digital signal processing while allowing simplified spectrum managing for complying to \ac{EMC} constraints. Originally considered for wired networks scenarios such as aircraft, a variation of the latter has been studied in \cite{parte2} considering a channel estimation procedure instead of the classical pulse compression.

For enabling more efficient sensing, an efficient solution is to perform distributed reflectometric and transferometric sensing of the monitored network. Aiming such purpose, research efforts have already been done for multicarrier-based \ac{TDR} systems \cite{lelong2010}. Also, multiple access schemes based on \ac{TDMA}, \ac{FDMA}, and \ac{CDMA}, have been studied for \ac{TDR} systems based on the basedband version of \ac{OFDM}, i.e. \ac{HS-OFDM}, in the context of power distribution network sensing \cite{parte2}.

Recently, \ac{OCDM}, which is a multicarrier scheme based on the \ac{DFnT}, has been proposed for wireless systems \cite{ouyang2016}. Such scheme has shown robustness to multipath propagation, being an interesting alternative for digital data communication. Furthermore, four modified versions of \ac{DFnT} that allow the extension of such scheme for baseband data transmission in \ac{PLC} systems have been proposed \cite{dib_dissert2018}. One of these four versions, named Type III \ac{OCDM} in \cite{dib_dissert2018}, adds Hermitian symmetry to the \ac{DFnT} matrix in order to produce a real-valued transmit vector in the discrete-time domain demanding low computational complexity.
Considering the modified \ac{DFnT} of the Type III \ac{OCDM} from \cite{dib_dissert2018}, this study describes an \ac{OCDM}-based \ac{TDR} system for power line sensing. Based on the convolution theorem of the \ac{DFnT}, which also holds for the modified \ac{DFnT}, a multiple access scheme is then introduced for the \ac{OCDM}-based \ac{TDR} system. Finally, the proposed multiple access scheme is compared with \ac{HS-OFDM}-based \ac{TDMA}, \ac{FDMA}, and \ac{CDMA} schemes in order to assess its effectiveness.

Given this background, the contributions of this study are as follows.
\begin{enumerate}
	\item Modelling of an \ac{OCDM}-based \ac{TDR} system for power line sensing considering the single \ac{PLM} case. In the context of this formulation, a multiple access scheme exploiting the convolution theorem of the modified \ac{DFnT} for distributed power line sensing is introduced.
	\item Performance assessment and comparison of the proposed \ac{OCDM}-based multiple access scheme for power line sensing via \ac{TDR} with the \ac{HS-OFDM}-based \ac{TDMA}, \ac{FDMA}, and \ac{CDMA} multiple access schemes.
\end{enumerate}
Based on the attained results, our major findings are as follows.
\begin{enumerate}
	\item The \ac{OCDM}-based multiple access scheme performs equally to the \ac{HS-OFDM}-based counterparts in terms of range resolution. Regarding maximum unambiguous range, it is more affected by the measurement window length in the proposed scheme and the \ac{HS-OFDM}/\ac{FDMA} scheme than in the remaining schemes respectively due to the subchirp and subcarrier allocation among the \acp{PLM}. The considered scenarios for assessing range resolution and maximum unambiguous range values attained by the proposed scheme are typical European underground low-voltage and US overhead medium-voltage power distribution networks, while the system parameters are based on \ac{NB}-\ac{PLC} standards.
	
	\item  Both the proposed \ac{OCDM}-based and the \ac{HS-OFDM}/\ac{FDMA} multiple access schemes yield much higher number of reflectograms and transferograms than the remaining considered schemes over time. They are, therefore, more appropriate for continuously performing distributed sensing of power lines and consequently sensible to intermittent faults, being the latter feature common to the \ac{HS-OFDM}/\ac{CDMA} scheme in certain cases \cite{lelong2010}.
	
	\item Considering a Brazilian overhead \ac{MV} scenario and maintaining the system parametrization, equal performance to the \ac{HS-OFDM}/\ac{TDMA} scheme in terms of \ac{SINR} is attained by the proposed \ac{OCDM}-based multiple access scheme. Meanwhile, the \ac{SINR} is increased or decreased at different \acp{PLM} in the \ac{HS-OFDM}/\ac{FDMA} scheme due to the combination of the subcarrier hopping among \acp{PLM}  and exponentially decaying \ac{PSD}. Finally, the best \ac{SINR} performance is attained by the \ac{HS-OFDM}/\ac{CDMA} scheme, which is a consequence of the noise averaging at the decoding process \cite{parte2,lelong2010}.
\end{enumerate}

The remainder of this paper is organized as follows. The \ac{OCDM}-based \ac{TDR} system for sensing of a power distribution network is described in Section~\ref{sec:model} considering the single \ac{PLM} case. Next, Section~\ref{sec:distributed} discusses range resolution and maximum unambiguous range aspects of the considered \ac{TDR} system and introduces a novel multiple access scheme based on \ac{OCDM}. Numerical results and discussions are presented in Section~\ref{sec:analysis} for evaluating the proposed multiple access scheme with \ac{HS-OFDM}-based counterparts, and concluding remarks are placed in Section~\ref{sec:conclusion}. Finally, the convolution theorem of the modified \ac{DFnT} is demonstrated in the Appendix.

\subsection*{Notation}\label{subsec:notation}

Throughout the paper, $\delta(t)$ and $\delta[k]$ are respectively the Dirac and Kronecker delta functions; $(\cdot)^\dagger$ indicates the Hermitian transpose operator; $\mathbb{E}\{\cdot\}$ represents the expectation operator; $\diag\{\cdot\}$ converts the argument vector into a diagonal matrix; $\mathbf{I}_{M}$ is a $M$-size identity matrix; the $M$-size \ac{DFT} matrix is denoted by $\mathbf{W}_{M}$; and $\mathbf{\Phi}_{M}$ is the $M$-size modified \ac{DFnT} matrix, which is the baseband version of the \ac{DFnT} matrix.

\section{System Model}\label{sec:model}

Let a baseband \ac{TDR} system be consisted by a full-duplex \ac{PLM} connected to a power distribution network, which, at a single given point, injects signals and captures reflections that travel at a phase velocity $v_p$. Assuming that the injection and subsequent capture of reflections of signals takes place within a coherence time $T_c$, in which variations in loads or any other element of the network are irrelevant, one can consider the power distribution network as \iac{LTI} system. The reflections captured by the \ac{PLM} from such power distribution network, which are raised by impedance discontinuities along the path traveled by the injected signal, are therefore the output of a reflection channel with impulse response expressed as
\begin{equation}\label{eq:h(t)}
h_\Gamma(t) = \sum_{i=0}^{N_\Gamma-1}\alpha_i\delta(t-T_i),
\end{equation}
where $N_\Gamma$ is the number of impedance discontinuities experienced by the injected signal, while $\alpha_i$ and $T_i$ are respectively the attenuation factor and the arrival time at the receiver side of the \ac{PLM} associated with the reflection raised by the $i^{th}$ impedance discontinuity, \mbox{$i\in\{0,\cdots,N_\Gamma-1\}$}.

A reflectometric sensing of the power distribution network, which consists of an analysis of the reflections raised by the impedance discontinuities along its length, can be used for applications such as topology inference and fault detection and location. Such task can be performed by a \ac{TDR} system which is capable of obtaining estimates of the reflection channel impulse response, named \textit{reflectograms}. In this study, it is assumed that such \ac{TDR} system is based on a baseband \ac{OCDM} scheme that is band-limited to a bandwidth $B$ and has sampling frequency $F_s=2B$.

\begin{figure*}[!t]
	\centering
	\psfrag{a}[c][c]{$\mathbf{\dot{x}}$}
	\psfrag{b}[c][c]{$\mathbf{s}$}
	\psfrag{c}[c][c]{$\mathbf{r}$}
	\psfrag{d}[c][c]{$\boldsymbol{\rho}$}
	\psfrag{P}[c][c]{$\mathcal{P}(\cdot)$}
	\psfrag{Q}[c][c]{$\mathcal{Q}(\cdot)$}
	\psfrag{s}[c][c]{$s(t)$}
	\psfrag{h}[c][c]{$h_\Gamma(t)$}
	\psfrag{r}[c][c]{$\tilde{r}(t)$}
	\psfrag{n}[c][c]{$v(t)$}
	\psfrag{y}[c][c]{$r(t)$}	
	\includegraphics[height=5cm]{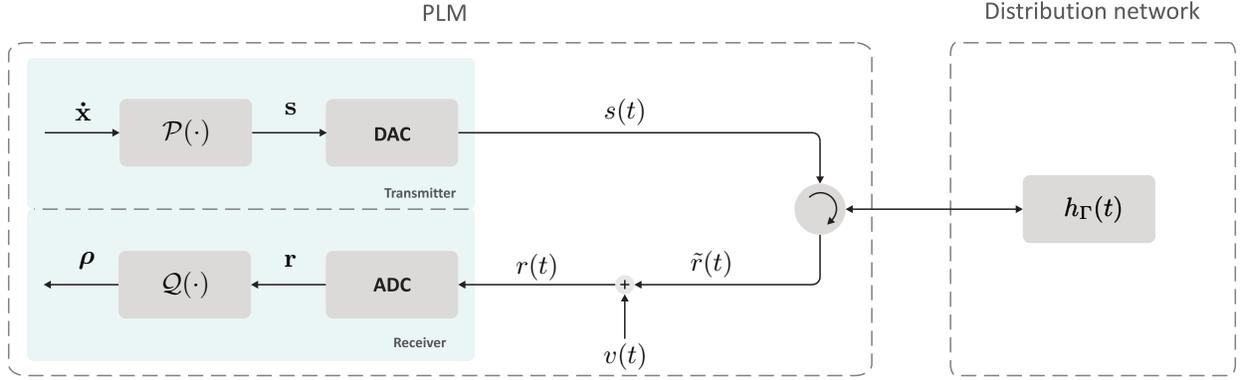}
	\caption{\ac{OCDM}-based baseband \ac{TDR} system over a distribution network.}\label{fig:OCDM_TDR}	
\end{figure*}

The considered \ac{OCDM}-based \ac{TDR} system is depicted in Fig.~\ref{fig:OCDM_TDR}, starting with a real discrete-Fresnel domain vector \mbox{$\mathbf{\dot{x}}=[\dot{x}_0,\dot{x}_1,\cdots,\dot{x}_{2N-1}]^T$}, such that $\mathbf{\dot{x}}\in\mathbb{R}^{2N\times1}$. This vector is inputted to the function $\mathcal{P}(\cdot)$ that represents, in a condensed form, the twofold digital processing performed at the transmitter side. The first processing stage performed by $\mathcal{P}(\cdot)$ is a modified \ac{IDFnT} on $\mathbf{\dot{x}}$, which generates the $2N$-length discrete-time domain vector $\mathbf{x} = \mathbf{\Phi}^\dagger_{2N}\mathbf{\dot{x}}$, $\mathbf{x}\in\mathbb{R}^{2N\times1}$. Finally, an $L_{cp}$-length cyclic prefix is appended to $\mathbf{x}$, resulting in the vector $\mathbf{s}\in\mathbb{R}^{(2N+L_{cp})\times1}$. Considering that the reflection channel impulse response has an $L_{h,\Gamma}$-length discrete-time domain representation \mbox{$\mathbf{h}_\Gamma=[h_{\Gamma,0},h_{\Gamma,1},\cdots,h_{\Gamma,L_{h,\Gamma}-1}]^T$}, $\mathbf{h}_\Gamma\in\mathbb{R}^{L_{h,\Gamma}\times1}$, no \ac{ISI} is experienced if the constraint $L_{cp}\leq L_{h,\Gamma}$ is satisfied.

Back to Fig.~\ref{fig:OCDM_TDR}, the discrete-time vector $\mathbf{s}$ undergoes an digital-to-analog conversion, being converted into the continuous-time signal $s(t)$ that is inputted to the reflection channel of impulse response $h_\Gamma(t)$. It is worth highlighting that, in order for the \ac{LTI} assumption of the reflection channel to hold, the duration of $s(t)$, i.e., the \ac{OCDM} symbol in the continuous-time domain, $T_{symb}=(2N+L_{cp})T_s$ must satisfy the constraint $T_{symb}\ll T_c$. The first term of the previous expression accounts for the number of samples of the discrete-time vector $\mathbf{s}$, while $T_s=1/F_s$ is the sampling period. The resulting signal from the convolution between the transmit signal and the channel is \mbox{$\tilde{r}(t)=s(t)\star h_\Gamma(t)$}. To this signal is added the noise $v(t)$, which is a zero-mean \ac{WSS} random process, resulting in the received signal \mbox{$r(t)=s(t)\star h_\Gamma(t)+v(t)$}. Note that, due to the noise presence, $r(t)$ is also a \ac{WSS} random process.

At the receiver side, the signal $r(t)$ passes through an analog-to-digital converter, originating the discrete-time domain vector $\mathbf{r}\in\mathbb{R}^{(2N+L_{cp})\times1}$. Next, $\mathbf{r}$ is inputted to the function $\mathcal{Q}(\cdot)$, which synthesizes the digital processing at the receiver side of the \ac{OCDM}-based \ac{TDR} system. The performed processing by this function starts with cyclic prefix removal from $\mathbf{r}$, which originates the discrete-time domain vector $\mathbf{y}\in\mathbb{R}^{2N\times1}$ expressed as
\begin{equation}\label{eq:circ_conv_time}
\mathbf{y}= \mathbf{H}_\Gamma \mathbf{x}+\mathbf{v},
\end{equation} 
where $\mathbf{H}_\Gamma\in\mathbb{R}^{2N\times2N}$ is a circulant matrix associated with a zero-padded version of $\mathbf{h}_\Gamma$, and $\mathbf{v}\in\mathbb{R}^{2N\times1}$ is a $2N$-length discrete-time domain window of the additive noise $v(t)$. 

The next processing stage of $\mathcal{Q}(\cdot)$ consists of performing a modified \ac{DFnT} on $\mathbf{y}$, originating the discrete-Fresnel domain vector $\mathbf{\dot{y}}\in\mathbb{R}^{2N\times1}$. As the convolution theorem of the \ac{DFnT} also holds for the modified \ac{DFnT}, which is demonstrated in the Appendix, $\mathbf{\dot{y}}$ can be equivalently expressed as
\begin{equation}\label{eq:circ_conv_Fresnel}
\mathbf{\dot{y}}=\mathbf{H}_\Gamma \mathbf{\dot{x}} + \mathbf{\dot{v}},
\end{equation} 
being \mbox{$\mathbf{\dot{v}}=[\dot{v}_0,\dot{v}_1,\cdots,\dot{v}_{2N-1}]^T$}, $\mathbf{\dot{v}}\in\mathbb{R}^{2N\times1}$, a $2N$-length discrete-Fresnel domain window of the additive noise $v(t)$.

The processing performed by $\mathcal{Q}(\cdot)$ is completed with the obtaining of an $L_\rho$-length reflectogram from $\mathbf{\dot{y}}$, denoted in the discrete-time domain as $\boldsymbol{\rho}\in\mathbb{R}^{L_\rho\times1}$, $L_\rho\geq L_{h,\Gamma}$. Besides ensuring high \ac{SNR} in order for noise effect to be negligible, a proper reflectogram must have the influence of the transmit vector $\mathbf{\dot{x}}$ on it minimized or eliminated so that it is a good estimate of the reflection channel impulse response $\mathbf{h}_{\Gamma}$. In this study, this is accomplished by simply designing a proper $\mathbf{\dot{x}}$ so that $\boldsymbol{\rho}$ is directly obtained from $\mathbf{\dot{y}}$ with no additional processing. Also, for covering cases where multiple \acp{PLM} are to perform network sensing simultaneously, this design must also enable orthogonal multiple access among the \acp{PLM}. A discussion on the design of $\mathbf{\dot{x}}$ is presented in Section~\ref{sec:distributed}.

Once $\boldsymbol{\rho}$ has been obtained, its analog counterpart $\rho(t)$ can be yielded in order to reduce the temporal granularity and, as a consequence, the spatial granularity of the reflectogram and the impedance discontinuity location accuracy. As discussed in \cite{parte2}, an interesting approach for the considered system would be sinc interpolation in the time domain by a reconstruction filter \cite{mitra2010} as the reflectogram is directly obtained in the discrete-time domain.

\section{Distributed Power Line Sensing}\label{sec:distributed}


As discussed in Section~\ref{sec:model}, for obtaining a proper reflectogram from the discrete-Fresnel domain vector $\mathbf{\dot{y}}$, the influence of the discrete-Fresnel domain transmit vector $\mathbf{\dot{x}}$ on it as expressed in \eqref{eq:circ_conv_Fresnel} must be minimized or ideally removed. Considering that a single \ac{PLM} performs sensing of the power distribution network, a transmit vector $\mathbf{\dot{x}}$ constituted of pilots spaced by $L_\rho$ samples in the discrete Fresnel domain can be used \cite{ouyang2017,ouyang2018}. Thus, the $k^{th}$ element of $\mathbf{\dot{x}}$ will be given by
\begin{equation}
\dot{x}_k = \sum\limits_{u=0}^{2N/L_{\rho} - 1} \delta[k - uL_{\rho}],
\end{equation}
in which \mbox{$k\in\{0,\cdots,2N-1\}$}. Based on the convolution theorem of the modified \ac{DFnT}, the $k^{th}$ element of the resulting discrete-Fresnel domain received vector $\mathbf{\dot{y}}$ can be expressed as
\begin{equation}
\dot{y}_k = \left(\sum\limits_{u=0}^{2N/L_{\rho} - 1} h_{\Gamma,k - uL_{\rho}}\right)+\dot{v}_k,
\end{equation}
Note that \mbox{$h_{\Gamma,k - uL_{\rho}}=0$} for \mbox{$k - uL_{\rho}<0$} and \mbox{$k - uL_{\rho}>L_{h,\Gamma}-1$}, as $h_{\Gamma,k}$ is only defined for \mbox{$k\in\{0,\cdots,L_{h,\Gamma}-1\}$}. As a consequence, $\mathbf{\dot{y}}$ will be constituted of concatenated estimates of the reflection channel impulse response impaired by additive noise, which will not cause mutual interference on each other if the spacing between consecutive pilots is equal to or longer than the length of $\mathbf{h}_\Gamma$, i.e. $L_\rho\geq L_{h,\Gamma}$. Finally, $2N/L_\rho$ unbiased discrete-time domain reflectograms are obtained by dividing the discrete-Fresnel domain vector $\mathbf{\dot{y}}$ into $L_\rho$-length windows. Such channel estimation is therefore equivalent to an ideal pulse compression procedure, whose capability of distinguishing reflections caused by close impedance discontinuities, i.e, range resolution, is given by \cite{parte1,parte2}
\begin{equation}
\Delta d = \frac{v_p}{4B}.
\end{equation}
Additionally, the \ac{PSLR} and \ac{ISLR} levels of each obtained estimate will only depend upon the occupied frequency bandwidth $B$ in the baseband by the \ac{TDR} system as described in \cite{parte2}.

In the proposed \ac{OCDM}-based \ac{TDR} system, the maximum unambiguous range is limited by both the measurement window length $L_\rho$ and the cyclic prefix length $L_{cp}$. As previously discussed, $L_\rho$ must be set so that there is no mutual interference among measurement windows. $L_{cp}$, in its turn, must be long enough so that no \ac{ISI} is experienced. Based on the discussion carried out in \cite{parte2}, the maximum unambiguous range for reflectograms is
\begin{equation}\label{eq:dmax_r}
d_{\max,R} = \frac{v_pT_s}{2}\mathop {\min \{L_\rho,L_{cp}\}}.
\end{equation}
For transferograms, the maximum unambiguous range is
\begin{equation}\label{eq:dmax_t}
d_{\max,T} = v_pT_s\mathop {\min \{L_\rho,L_{cp}\}},
\end{equation}
which is twice the value assumed by $d_{\max,r}$, in which the round trip time for the transmitted travelling waves is considered.

Moving to the case where $N_{PLM}$ \acp{PLM} perform sensing of a single power distribution network simultaneously, an efficient multiple access strategy must be adopted. In the case of \ac{HS-OFDM}-based \ac{TDR} systems, \ac{TDMA}, \ac{FDMA}, and \ac{CDMA} schemes have been discussed in a previous study \cite{parte2}. For the case of the \ac{OCDM}-based \ac{TDR} system considered in this study, the orthogonality among the signals associated with the multiple \acp{PLM} is ensured by a proper subchirp allocation among them.

\begin{figure}[!t]
	\centering
	\psfrag{B}[c][c]{\small $L_\rho$}
	\psfrag{C}[c][c]{\footnotesize $0$}
	\psfrag{D}[c][c]{\footnotesize $u$}
	\psfrag{E}[c][c]{\tiny $N_{\scalebox{.85}{PLM}}-1$}
	\psfrag{F}[c][c]{\footnotesize $0$}
	\psfrag{G}[c][c]{\footnotesize $1$}
	\psfrag{H}[c][c]{\tiny $N_{\scalebox{.85}{PLM}}-1$}
	\includegraphics[width=8.5cm]{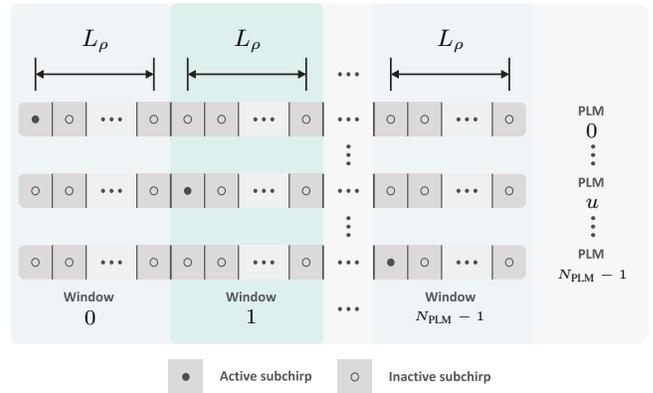}
	\caption{Subchirp allocation among the \acp{PLM}.}\label{fig:chirp_allocation}	
\end{figure}

The proposed subchirp allocation scheme for enabling multiple access among \acp{PLM} performing unbiased power line sensing is shown in Fig.~\ref{fig:chirp_allocation}. In this figure, the subchirp pilot of index $k = uL_\rho$ of the $2N$-length discrete-Fresnel domain transmit vector $\mathbf{\dot{x}}$ is active for the $u^{th}$ \ac{PLM}, \mbox{$u\in\{0,\cdots,N_{PLM}-1\}$}. For the sake of fairness, the distance between the sample associated with the active pilot of the $u^{th}$ \ac{PLM} and sample associated with the active pilot of the $(u+1)^{th}$ \ac{PLM} is set to $L_\rho=2N/N_{PLM}$. 

At the receiver side, the vector $\mathbf{\dot{y}}$ at the $u^{th}$ \ac{PLM} will be composed by $N_{PLM}$ measurement windows. The $u^{th}$ measurement window starts at the subchirp $k=uL_\rho$ and ends at the subchirp $k=uL_\rho+L_\rho-1$ of $\mathbf{\dot{y}}$ and it is a reflectogram at the $u^{th}$ \ac{PLM} and a transferogram at the remaining \acp{PLM}. In order to perform effective network sensing, data fusion techniques can be applied on the obtained measurement windows \cite{auzanneau2016}.

In order to avoid interference among measurement windows, the spacing among pilots must satisfy the constraint \mbox{$L_\rho=2N/N_{PLM}\geq L_{h,\max}$}, or alternatively
\begin{equation}
	2N\geq N_{PLM}L_{h,\max}.
\end{equation}
For this constraint, $L_{h,\max}$ is defined as
\begin{equation}
	L_{h,\max} \triangleq \max\limits_{ij}\{L_{h,ij}\},
\end{equation}
where $L_{h,ij}$ is the length of the channel impulse response $\mathbf{h}_{ij}\in\mathbb{R}^{L_{h,ij}\times1}$ for a signal injected by the $j^{th}$ \ac{PLM} into the network and received by the $i^{th}$ \ac{PLM}, with \mbox{$i\in\{0,\cdots,N_{PLM}-1\}$} and \mbox{$j\in\{0,\cdots,N_{PLM}-1\}$}. Altogether, there are $N^2_{PLM}$ channel impulse responses $\mathbf{h}_{ij}$, with a total of $N_{PLM}$ reflectograms ($\mathbf{h}_{ij}$, $i=j$) and \mbox{$N_{PLM}(N_{PLM}-1)$} transferograms ($\mathbf{h}_{ij}$, $i\neq j$).

The proposed multiple access scheme and system parametrization result in the obtaining of a number
\begin{equation}\label{eq:N_rho}
N_{\rho}=\frac{1}{T_{symb}}
\end{equation}
reflectograms per \ac{PLM} per second. Additionally, a number is
\begin{equation}\label{eq:N_tau}
N_{\tau}=\frac{N_{PLM}-1}{T_{symb}}
\end{equation}
of transferograms are obtained per \ac{PLM} per second,
resulting in a total number of \mbox{$N_{meas}=N_{\rho}+N_{\tau}=N_{PLM}/T_{symb}$} measurements per \ac{PLM} per second.

\section{Numerical Analysis}\label{sec:analysis}

In order to validate the carried out discussion, numerical results are presented and discussed in this section.
Given this context, Subsection~\ref{subsec:rr_mur} presents range resolution and maximum unambiguous range values for the \mbox{\ac{OCDM}-based} \ac{TDR} system considering typical system parametrizations for \mbox{\ac{NB}-\ac{PLC}} systems, while Subsection~\ref{subsec:MA} carries out a comparative analysis of the proposed \mbox{\ac{OCDM}-based} distributed \ac{TDR} system with the \mbox{\ac{HS-OFDM}-based} counterpart with \ac{TDMA}, \ac{FDMA}, and \ac{CDMA} multiple access schemes described in a previous study \cite{parte2}.

\subsection{Range resolution and maximum unambiguous range}\label{subsec:rr_mur}

For performing proper sensing of a power distribution network, it is paramount to have awareness of the \ac{TDR} system limitations resulting from its parametrization. In this context, an European underground low-voltage power distribution network and an US overhead medium-voltage power distribution network in a rural area are considered for range resolution and maximum unambiguous range analysis. For the low-voltage scenario, it is considered a power supply cable NAYY150SE with resistance $R'$, inductance $L'$, conductance $G'$, and capacitance $C'$ per unit length calculated as in \cite{lampe_vinck2011}, whereas for the medium-voltage cable, the power supply cable with distributed parameters listed in \cite{1901_2} is adopted.  Based on these parameters, the phase velocity is calculated by $v_p=1/\sqrt{L'C'}$ \cite{paul2007}, resulting in $v_p=1.50\times10^8$ for the considered \ac{LV} cable, and $v_p=2.56\times10^8$ for the considered \ac{MV} cable.

\begin{figure}[!b]
	\centering
	\psfrag{XX}[c][c]{$B$(kHz)}
	\psfrag{YY}[c][c]{$\Delta d$(m)}
	\includegraphics[width=8.5cm]{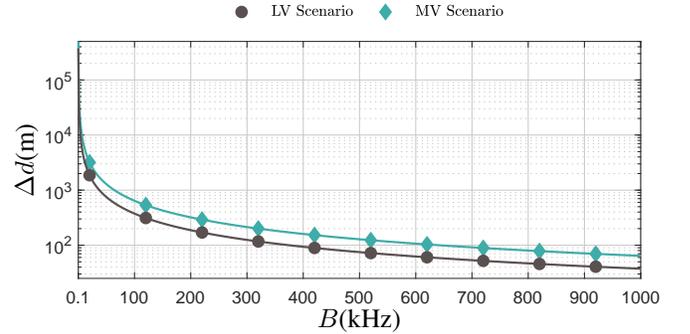}
	\caption{Range resolution $\Delta d$ in meters as a function of the occupied frequency bandwidth $B$ in the considered LV and MV scenarios \cite{parte2}.}\label{fig:RR}
\end{figure}

Fig.~\ref{fig:RR} shows the range resolution $\Delta d$ as a function of the occupied frequency bandwidth $B$ for the considered \ac{LV} and \ac{MV} scenarios.  The achieved $\Delta d$ values range from tens of thousands of kilometers for low $B$ values to a few meters for higher $B$ values, with a ratio of $1.71$ between the resolution in the \ac{MV} and \ac{LV} and  scenarios due to their different phase velocity. The presented results indicate that $B$ values in the \ac{NB}-\ac{PLC} frequency range, i.e., $B<500$~kHz, result in fair range resolution values, i.e., $\Delta d\geq75$~m, and therefore a fair capability of resolving close impedance discontinuities for typical distances covered by \ac{PLC} signaling in \ac{LV} and \ac{MV} power distribution networks, which are about $1$~km in \ac{MV} scenarios and shorter in \ac{LV} scenarios \cite{passerini2018_2,PLCbook2016}. Given this result, the parameters listed in Table~\ref{tab:parameters}, which are based on typical \ac{NB}-\ac{PLC} standards \cite{1901_2}, are adopted for the results presented henceforth.

\begin{figure}[!b]
	\centering
	\psfrag{XX}[c][c]{$N_{PLM}$}
	\psfrag{YY}[c][c]{$d_{\max,\rho}$(km)}
	\psfrag{ZZ}[c][c]{$d_{\max,\tau}$(km)}
	\hspace*{-5.3mm}
	\includegraphics[width=9.7cm]{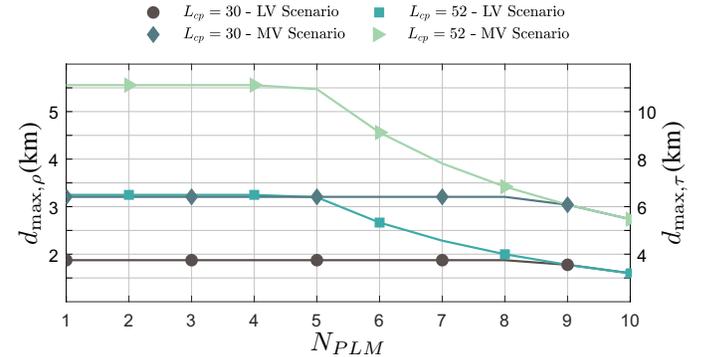}
	\caption{Maximum unambiguous range for reflectograms $d_{\max,\rho}$ and transferograms $d_{\max,\tau}$ as functions of the number of \acp{PLM} $N_{PLM}$.}\label{fig:MUR}
\end{figure}

Next, Fig.~\ref{fig:MUR} shows the maximum unambiguous ranges $d_{\max,\rho}$ and $d_{\max,\tau}$ as functions of the cyclic prefix length $L_{cp}$ for different numbers of \acp{PLM} $N_{PLM}$ in the considered \ac{LV} and \ac{MV} scenarios. One observes \mbox{$d_{\max,\rho}=1.87$~km} and \mbox{$d_{\max,\tau}=3.75$~km} in the \ac{LV} scenario and \mbox{$d_{\max,\rho}=3.21$~km} and \mbox{$d_{\max,\tau}=6.41$~km} in the \ac{MV} scenario for $N_{PLM}\leq4$ if $L_{cp}=30$. For $N_{PLM}>4$, the measurement window length $L_{\rho}$ becomes shorter than $L_{cp}$, reducing the maximum unambiguous ranges according to the relation described in \eqref{eq:dmax_r} and \eqref{eq:dmax_t}. In the case where $L_{cp}=52$ is adopted, a similar behavior is observed. For $N_{PLM}\leq8$, maximum unambiguous range values of \mbox{$d_{\max,\rho}=3.25$~km} and \mbox{$d_{\max,\tau}=6.50$~km} are observed in the \ac{LV} scenario, while the values \mbox{$d_{\max,\rho}=5.56$~km} and \mbox{$d_{\max,\tau}=11.11$~km} are observed in the \ac{MV} scenario. As in the previous case, an increase in $N_{PLM}$ results in considerable reduction of the maximum unambiguous range values.

\begin{table}[!t]
	\renewcommand{\arraystretch}{1.5}
	\arrayrulecolor[HTML]{708090}
	\setlength{\arrayrulewidth}{.1mm}
	\setlength{\tabcolsep}{4pt}
	
	\footnotesize
	\centering
	\caption{Adopted system parameters.}
	\label{tab:parameters}
	\begin{tabular}{c|c|c|c}
		\hline\hline
		\multicolumn{4}{c}{\textbf{System parameters}}\\
		\hline\hline
		\multicolumn{3}{c|}{\textbf{Frequency range (kHz)}}                    & $0-500$\\ \hline
		\multicolumn{3}{c|}{\textbf{Frequency bandwidth} $B$~\textbf{(kHz)}}   & $500$\\ \hline
		\multicolumn{3}{c|}{\textbf{Sampling frequency} $F_s$~\textbf{(MHz)}}  & $1$\\ \hline
		\multicolumn{3}{c|}{\textbf{OCDM symbol length} $2N$}                  & $256$\\ \hline
		\multirow{2}{*}{\textbf{Cyclic prefix length} $L_{cp}$}                & \multicolumn{2}{c|}{Standard}  & $30$ \\
		& \multicolumn{2}{c|}{Long}      & $52$ \\ \hline\hline
	\end{tabular}
\end{table}

\subsection{Comparison with multiple access schemes for HS-OFDM-based \ac{TDR} systems}\label{subsec:MA}

\begin{figure*}[!b]
	\centering	
	\psfrag{A}[c][c]{\footnotesize $d_a=1$~km}
	\psfrag{B}[c][c]{\footnotesize $d_b=1.73$~km}
	\psfrag{T}[c][c]{MV}
	\includegraphics[height=4.5cm]{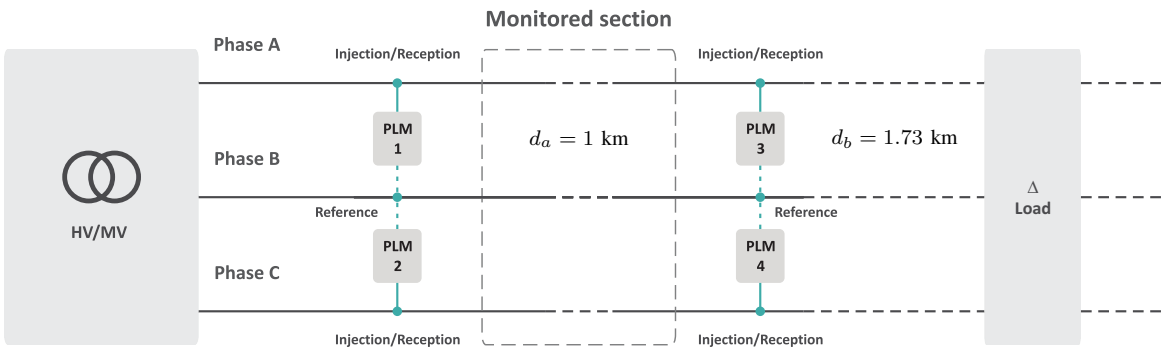}
	\vspace{0.5cm}
	\caption{Considered MV distribution network section \cite{parte2}.}\label{fig:MV_line}	
\end{figure*}

The next aspect of the numerical analysis is a comparison of the proposed \ac{OCDM}-based multiple access scheme for distributed power line sensing with the \ac{TDMA}, \ac{FDMA}, and \ac{CDMA} multiple access schemes described in \cite{parte2} for \ac{HS-OFDM}-based \ac{TDR} systems. For this purpose, the system parameters listed in \ref{tab:parameters} are maintained and \ac{BPSK} modulation is adopted. Additionally, an \ac{MTL}-based model \cite{paul2007,franek2017} of a real \ac{MV} power distribution network section in the city of Curitiba, Brazil,  described in \cite{parte2} and depicted in Fig.~\ref{fig:MV_line} is considered. Finally, the adopted additive noise model is the one reported in \cite{tao2007,girotto2017}, whose one-sided \ac{PSD} in the continuous-frequency domain is $S_V(f)=-93+52.98e^{-0.0032f/10^3}$~dBm/Hz.

\begin{figure}[!t]
	\centering
	\subfloat[ ]{
		\psfrag{XX}[c][c]{$N_{PLM}$}
		\psfrag{YY}[c][c]{\small $N_{\rho}\left(\text{s}^{-1}\right)/10^{3}$}
		\includegraphics[width=4.5cm]{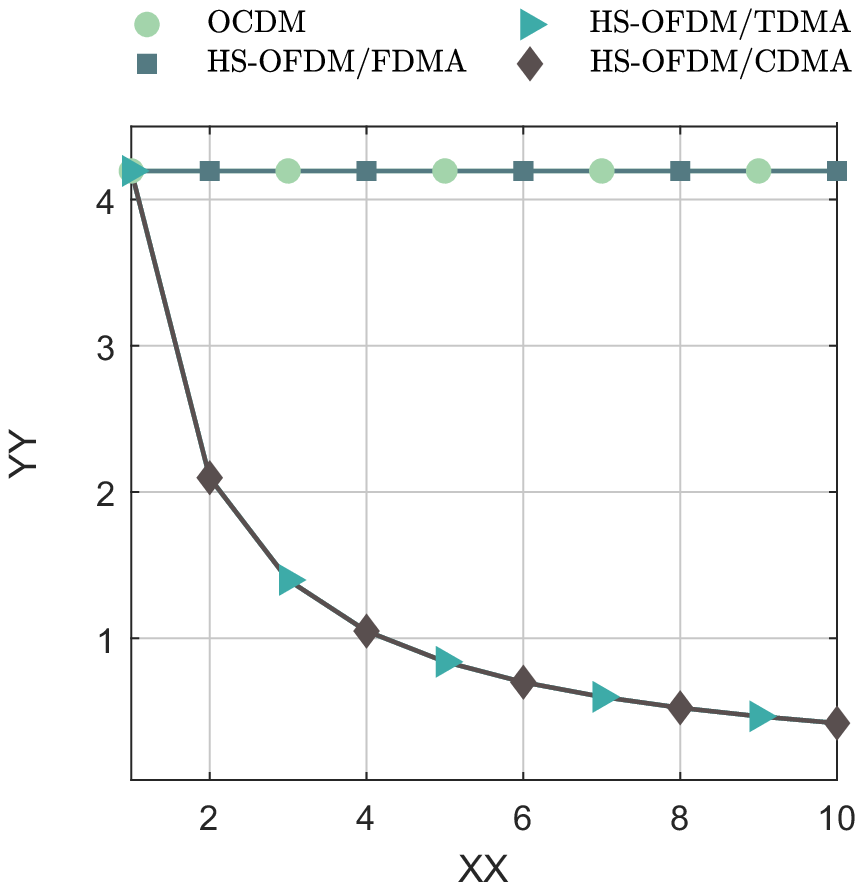}\label{fig:N_rho_st}
	}
	\subfloat[ ]{
		\psfrag{XX}[c][c]{$N_{PLM}$}
		\psfrag{YY}[c][c]{\small $N_{\rho}\left(\text{s}^{-1}\right)/10^{3}$}\label{fig:N_rho_long}
		\includegraphics[width=4.5cm]{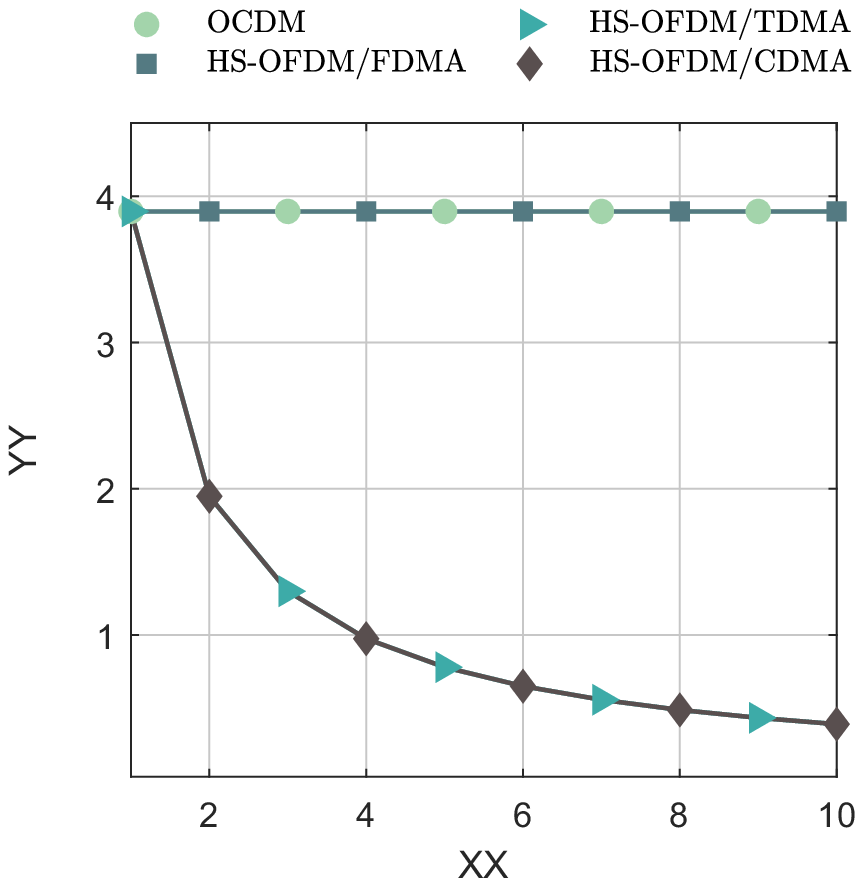}
	}
	\caption{Number of obtained reflectograms $N_{\rho}$ as a function of the number of \acp{PLM} $N_{PLM}$ for (a) $L_{cp}=30$, and (b) $L_{cp}=52$.}
\end{figure}

Given the considered scenario, the number of obtained reflectograms over time $N_\rho$ for the considered multiple access schemes as a function of $N_{PLM}$ is shown in Fig.~\ref{fig:N_rho_st} for $L_{cp}=30$ and in Fig.~\ref{fig:N_rho_long} for $L_{cp}=52$. While \ac{TDMA} and \ac{CDMA} for \ac{HS-OFDM}-based \ac{TDR} systems yield a decreasing number of reflectograms along with the number of \acp{PLM} due to time multiplexing and \ac{HS-OFDM} symbol spreading, respectively, the use of the proposed \ac{OCDM}-based multiple access scheme, as well as of the \ac{HS-OFDM}/\ac{FDMA}, results in a higher and constant number of reflectograms regardless of the number of \acp{PLM}.

\begin{figure}[!t]
	\centering
	\subfloat[ ]{
		\psfrag{XX}[c][c]{$N_{PLM}$}
		\psfrag{YY}[c][c]{\small $N_{\tau}\left(\text{s}^{-1}\right)/10^{3}$}
		\includegraphics[width=4.6cm]{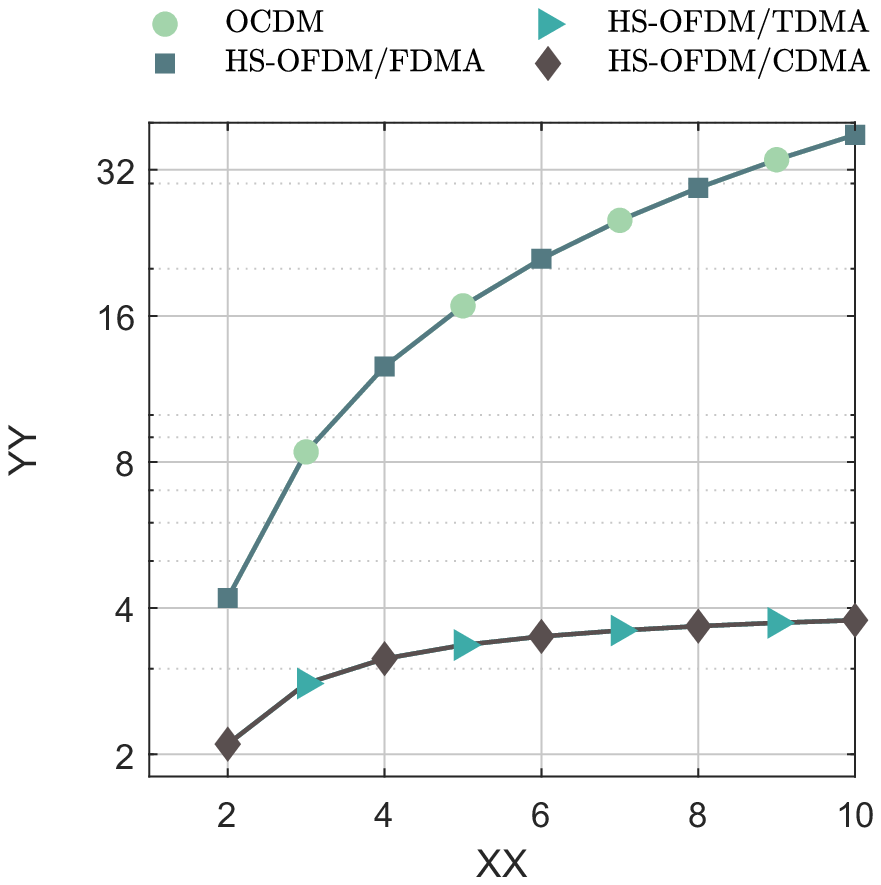}\label{fig:N_tau_st}
	}
	\subfloat[ ]{
		\psfrag{XX}[c][c]{$N_{PLM}$}
		\psfrag{YY}[c][c]{\small $N_{\tau}\left(\text{s}^{-1}\right)/10^{3}$}\label{fig:N_tau_long}
		\includegraphics[width=4.6cm]{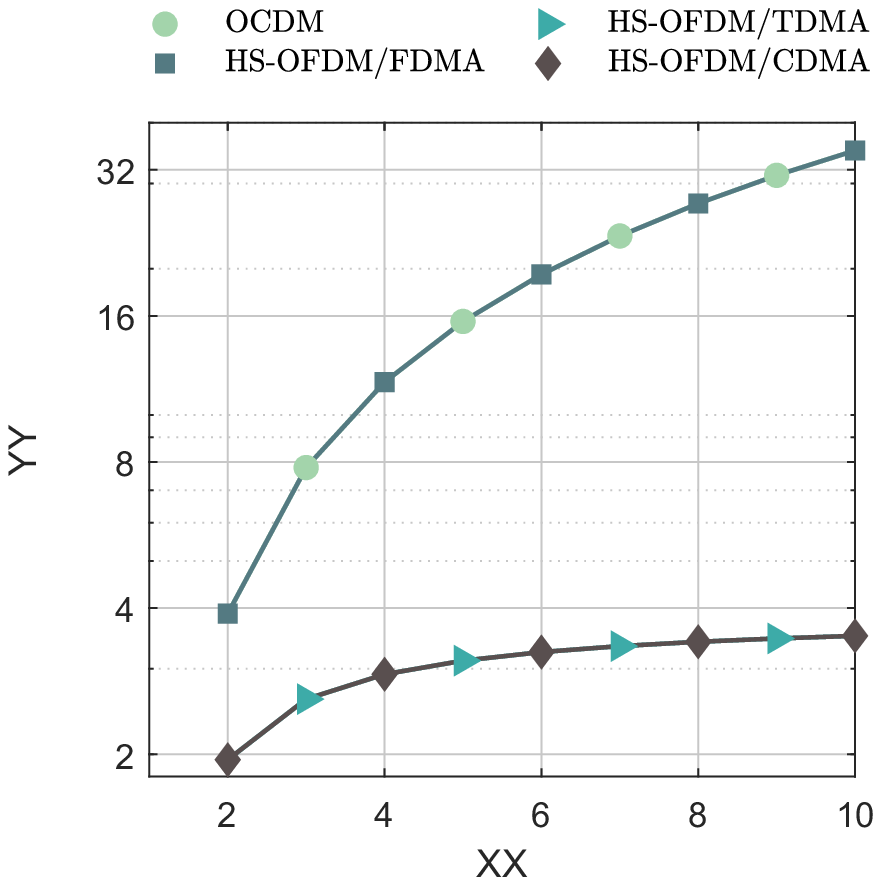}
	}
	\caption{Number of obtained transferograms $N_{\tau}$ as a function of the number of \acp{PLM} $N_{PLM}$ for (a) $L_{cp}=30$, and (b) $L_{cp}=52$.}
\end{figure}

The attained number of transferograms over time $N_\tau$ is shown Fig.~\ref{fig:N_tau_st} for $L_{cp}=30$ and in Fig.~\ref{fig:N_tau_long} for $L_{cp}=52$. As expected according to \eqref{eq:N_tau}, no transferogram is obtained for $N_{PLM}=1$. The presented results also show rapidly increasing $N_{\tau}$ values along with $N_{PLM}$ for the proposed \ac{OCDM}-based scheme and \ac{HS-OFDM}/\ac{FDMA}, while \ac{HS-OFDM}/\ac{TDMA} and \ac{HS-OFDM}/\ac{CDMA} yield slower increasing $N_{\tau}$ along with $N_{PLM}$. The reasons for such behavior are the same as in the case of reflectograms.

The last stage of the numerical analysis of this study is a performance comparison among the proposed \ac{OCDM}-based multiple access scheme and the ones discussed for \ac{HS-OFDM}-based \ac{TDR} systems in terms of the associated \ac{SINR} to the obtained measurements. For the sake of simplicity, only reflectograms are considered. Also, an one-sided \ac{PSD} of $-40$~dBm/Hz for the transmit signal has been adopted. Such assumption results in total transmission power values of $20$~dBm for the proposed \ac{OCDM}-based multiple access scheme as well as the \ac{HS-OFDM}/\ac{TDMA} and \ac{HS-OFDM}/\ac{CDMA} schemes, while the total transmission power for the \ac{HS-OFDM}/\ac{FDMA} scheme is $13.98$~dBm as not all subcarrier are active at each \ac{PLM}. Of all the considered schemes, only the \ac{HS-OFDM}/\ac{CDMA} presents mutual interference among the \acp{PLM} as described in \cite{parte2}, being the \ac{SINR} for the \ac{OCDM}-based scheme as well as for \ac{HS-OFDM}/\ac{TDMA} and \ac{HS-OFDM}/\ac{FDMA} equal to their associated \ac{SNR}.

The resulting \ac{SINR} for the four \acp{PLM} in the considered multiple access schemes are presented in Table~\ref{tab:SINR}. Due to the exponentially decreasing additive noise \ac{PSD} and the subcarrier hopping, the \ac{HS-OFDM}/\ac{FDMA} yields higher \ac{SINR} values at \acp{PLM} to which higher-frequency subcarriers are allocated. In the \ac{HS-OFDM}/\ac{CDMA} scheme, the higher \ac{SINR} values, which unlike the \ac{FDMA} case are fair among the \ac{PLM}, are explained by the averaging process in the decoding procedure \cite{parte2,lelong2010}. Finally, \ac{HS-OFDM}/\ac{TDMA} and the proposed \ac{OCDM}-based multiple access scheme yield fair \ac{SINR} values without any increase or decrease, and at a same level for all \acp{PLM}. This happens due to two factors, which are: i) unlike the \ac{HS-OFDM}/\ac{FDMA} case, all spectral content in the frequency bandwidth $B$ is contained in the reflectogram; and ii) there is no averaging process reducing the effective noise power and therefore increasing the \ac{SINR} as in the \ac{HS-OFDM}/\ac{CDMA} scheme.

\section{Conclusion}\label{sec:conclusion}

In this study, a novel multiple access scheme for \ac{OCDM}-based \ac{TDR} systems performing distributed power line sensing has been proposed. First, an \ac{OCDM}-based \ac{TDR} system considering a single \ac{PLM} has been described, being its range resolution and maximum unambiguous range aspects for both reflectograms and transferograms addressed. Next, the proposed multiple access scheme has been introduced and its implications on system parametrization and performance have been discussed. Finally, a numerical analysis has been carried out to support the discussion carried out throughout the paper.

\begin{table}[!t]
	\renewcommand{\arraystretch}{1.5}
	\arrayrulecolor[HTML]{708090}
	\setlength{\arrayrulewidth}{.1mm}
	\setlength{\tabcolsep}{4pt}
	
	\footnotesize
	\centering
	\caption{\ac{SINR} of the reflectograms at the four \acp{PLM} for the considered multiple access schemes.}
	\label{tab:SINR}
	\begin{tabular}{c|cccc}
		\hline\hline
		\multicolumn{5}{c}{\textbf{SINR~(dB)}}\\
		\hline\hline
		\textbf{PLM index} $u$  & 1           & 2           & 3           & 4       \\ \hline
		\textbf{OCDM} 			& $14.85$dB & $14.85$dB & $14.85$dB & $14.85$dB \\ \hline
		\textbf{HS-OFDM/TDMA}   & $14.85$dB & $14.85$dB & $14.85$dB & $14.85$dB \\ \hline
		\textbf{HS-OFDM/FDMA}   & $13.83$dB & $14.54$dB & $15.28$dB & $16.04$dB \\ \hline
		\textbf{HS-OFDM/CDMA}   & $17.87$dB & $17.87$dB & $17.87$dB & $17.87$dB \\ \hline\hline
	\end{tabular}
\end{table}

The attained results have illustrated the dependence of the range resolution on the occupied frequency bandwidth as well as the effect of measurement window length and cyclic prefix length on the maximum unambiguous range, both considering typical European underground \ac{LV} and US overhead \ac{MV} scenarios. A comparison with a previous work \cite{parte2} shows that such effects are equally present in the \ac{HS-OFDM}-based \ac{TDR} systems based on \ac{TDMA}, \ac{FDMA}, and \ac{CDMA} multiple access schemes, being the range resolution equal among all schemes and the maximum unambiguous range the most influenced by the measurement window length in the proposed \ac{OCDM}-based and \ac{HS-OFDM}/\ac{FDMA} schemes respectively due to the subchirp and subcarrier allocation among the \acp{PLM}. On the other hand, these schemes yield a much higher number of reflectograms and transferograms over time than the \ac{HS-OFDM}/\ac{TDMA} and \ac{HS-OFDM}/\ac{CDMA} schemes, which are respectively impaired by their time multiplexing and spreading processes. The proposed \ac{OCDM}-based multiple access scheme is therefore appropriate for continuously performing distributed sensing of a power line.

Finally, the results obtained considering the described Brazilian overhead \ac{MV} scenario have shown that the proposed scheme performs equally to the \ac{HS-OFDM}/\ac{TDMA} scheme in terms of \ac{SINR}. In comparison to the \ac{HS-OFDM}/\ac{FDMA} scheme, the \ac{OCDM}-based scheme performs better for 2 \ac{PLM} and worse for the other 2 \ac{PLM} due to the combined effect of subcarrier hopping and exponentially decaying \ac{PSD} that increases or decreases the \ac{SINR} for different \acp{PLM} in the former scheme. The best performance, however, is still attained by the \ac{HS-OFDM}/\ac{CDMA} scheme due to its averaging process that reduces the effective additive noise power.

\section*{Appendix}

To introduce the modified \ac{DFnT}, the diagonal matrix $\mathbf{\Gamma}_{2N}=\diag\{[\Gamma_{0,0},\Gamma_{1,1},\cdots,\Gamma_{2N-1,2N-1}]\}$ is defined. The elements $\Gamma_{k,k}$, $k=0,\cdots,2N-1$ of $\mathbf{\Gamma}_{2N}\in\mathbb{C}^{2N\times2N}$ are defined as
\begin{equation}\label{eq:eigenval_kk}
\Gamma_{k,k} \triangleq \left\{\arraycolsep=3pt\def\arraystretch{1.3}
\begin{array}{ll}
e^{(-j\pi/2N)(k^2)}, & k\leq N\\
e^{(j\pi/2N)(k^2)}, & k>N
\end{array}\right. .
\end{equation}
It clearly holds that $\mathbf{\Gamma}_{2N}\mathbf{\Gamma}^\dagger_{2N}=\mathbf{I}_{2N}$.
The diagonal matrix $\mathbf{\Gamma}_{2N}$ inputs are the eigenvalues of a circulant matrix $\mathbf{\Phi}_{2N}\in\mathbb{R}^{2N\times2N}$ such that 
\begin{equation}\label{eq:DFnT_diag}
\mathbf{F}_{2N}\mathbf{\Phi}_{2N}\mathbf{F}^\dagger_{2N}=\mathbf{\Gamma}_{2N},
\end{equation}
where $\mathbf{F}_{2N} = (1/\sqrt{2N})\mathbf{W}_{2N}$. The real circulant matrix $\mathbf{\Phi}_{2N}$ from \eqref{eq:DFnT_diag} is the modified \ac{DFnT} matrix used for transforming a discrete-time domain vector $\mathbf{c}\in\mathbb{R}^{2N\times2N}$ into the discrete-Fresnel domain vector $\mathbf{\dot{c}}\in\mathbb{R}^{2N\times2N}$ via the operation
\begin{equation}
	\mathbf{\dot{c}} = \mathbf{\Phi}_{2N}\mathbf{c},
\end{equation}
The inverse \ac{DFnT} is achieved therefore by
\begin{equation}
 \mathbf{c} = \mathbf{\Phi}^\dagger_{2N}\mathbf{\dot{c}}.
\end{equation}

The convolution theorem, as well as the unitary and the similarity transformation properties of the modified \ac{DFnT}, on which the former is based, are demonstrated as follows.

\subsection*{Unitary property of the modified \ac{DFnT}}

For the circulant modified \ac{DFnT} matrix $\mathbf{\Phi}_{2N}$ from \eqref{eq:DFnT_diag}, it holds the unitary property, i.e.,
\begin{eqnarray}\label{eq:unitary}
	\mathbf{\Phi}^\dagger_{2N}\mathbf{\Phi}_{2N} & = & (\mathbf{F}^\dagger_{2N}\mathbf{\Gamma}_{2N}\mathbf{F}_{2N}\nonumber)(\mathbf{F}^\dagger_{2N}\mathbf{\Gamma}^\dagger_{2N}\mathbf{F}_{2N})\\
	& = & \mathbf{I}_{2N}.
\end{eqnarray}
Due to the Hermitian symmetry of $\mathbf{\Gamma}_{2N}$ observed in \eqref{eq:eigenval_kk}, one has that $\mathbf{\Phi}_{2N}\in\mathbb{R}^{2N\times2N}$, which enables baseband applications for the presented \ac{DFnT} matrix.

\subsection*{Similarity transformation of the modified \ac{DFnT}}

For the given circulant \ac{DFnT} matrix, it also holds the similarity transformation expressed as
\begin{equation}\label{eq:similarity}
	\mathbf{\Phi}_{2N}\mathbf{Z}\mathbf{\Phi}^\dagger_{2N} = \mathbf{Z},
\end{equation}
in which $\mathbf{Z}\in\mathbb{R}^{2N\times2N}$ is a circulant matrix.
The proof for \eqref{eq:similarity2} starts by rearranging the expression as
\begin{equation}\label{eq:similarity2}
\mathbf{\Phi}_{2N}\mathbf{Z}\mathbf{\Phi}^\dagger_{2N} = (\mathbf{F}^\dagger_{2N}\mathbf{\Gamma}^\dagger_{2N}\mathbf{F}_{2N})\mathbf{Z}(\mathbf{F}^\dagger_{2N}\mathbf{\Gamma}_{2N}\mathbf{F}_{2N}).
\end{equation}
We next write
\begin{equation}\label{eq:similarity3}
\mathbf{\Phi}_{2N}\mathbf{Z}\mathbf{\Phi}^\dagger_{2N} = \mathbf{F}^\dagger_{2N}\mathbf{\Gamma}^\dagger_{2N}\mathbf{\Gamma}_Z\mathbf{\Gamma}_{2N}\mathbf{F}_{2N},
\end{equation}
in which $\mathbf{\Gamma}_Z=\mathbf{F}_{2N}\mathbf{Z}\mathbf{F}^\dagger_{2N}$ is a diagonal matrix whose inputs are the eigenvalues of the circulant matrix $\mathbf{Z}$. The pre- and post-multiplication of $\mathbf{\Gamma}_Z$ respectively by the diagonal matrices $\mathbf{\Gamma}^\dagger_{2N}$ and $\mathbf{\Gamma}_{2N}$ yields $\mathbf{\Gamma}^\dagger_{2N}\mathbf{\Gamma}_Z\mathbf{\Gamma}_{2N}$, which, as the three matrices are diagonal, becomes $\mathbf{\Gamma}^\dagger_{2N}\mathbf{\Gamma}_{2N}\mathbf{\Gamma}_Z=\mathbf{\Gamma}_Z$. We then rewrite \eqref{eq:similarity3} as
\begin{eqnarray}\label{eq:similarity4}
\mathbf{\Phi}_{2N}\mathbf{Z}\mathbf{\Phi}^\dagger_{2N} & = & \mathbf{F}^\dagger_{2N}\mathbf{\Gamma}_Z\mathbf{F}_{2N}\nonumber\\
& = & \mathbf{F}^\dagger_{2N}(\mathbf{F}_{2N}\mathbf{Z}\mathbf{F}^\dagger_{2N})\mathbf{F}_{2N}\nonumber\\
& = & \mathbf{Z},
\end{eqnarray}
which proves \eqref{eq:similarity}.

\subsection*{Convolution theorem of the modified \ac{DFnT}}

Considering two vectors $\mathbf{a}\in\mathbb{R}^{2N\times1}$ and $\mathbf{b}\in\mathbb{R}^{2N\times1}$, to which are associated the circulant matrices $\mathbf{A}\in\mathbb{R}^{2N\times2N}$ and $\mathbf{B}\in\mathbb{R}^{2N\times2N}$, respectively, one can write the circular convolution between $\mathbf{a}$ and $\mathbf{b}$ as
\begin{eqnarray}\label{eq:c}
	\mathbf{c} & = & \mathbf{A}\mathbf{b}\\
	& = & \mathbf{B}\mathbf{a},
\end{eqnarray}
where $\mathbf{c}\in\mathbb{R}^{2N\times1}$. By applying the \ac{DFnT} on $\mathbf{c}$, one obtains its discrete Fresnel domain representation $\mathbf{\dot{c}}\in\mathbb{R}^{2N\times1} = \mathbf{\Phi}_{2N}\mathbf{c}$, which can be rewritten as
\begin{eqnarray}
	\mathbf{\dot{c}} & = & \mathbf{\Phi}_{2N}\mathbf{A}\mathbf{b}\label{eq:c_fresnel1}\\
	& = & \mathbf{\Phi}_{2N}\mathbf{B}\mathbf{a}\label{eq:c_fresnel2}.
\end{eqnarray}
Making use of the unitary and similarity transformation properties of the \ac{DFnT} from \eqref{eq:unitary} and \eqref{eq:similarity}, respectively, \eqref{eq:c_fresnel1} and \eqref{eq:c_fresnel2} can be further rearranged respectively as
\begin{eqnarray}
\mathbf{\dot{c}} & = & \mathbf{\Phi}_{2N}\mathbf{A}\mathbf{\Phi}^\dagger_{2N}\mathbf{\Phi}_{2N}\mathbf{b}\label{eq:c_fresnel3}\\
& = & \mathbf{\Phi}_{2N}\mathbf{B}\mathbf{\Phi}^\dagger_{2N}\mathbf{\Phi}_{2N}\mathbf{a}\label{eq:c_fresnel4}.
\end{eqnarray}
Using the similarity transformation property from \eqref{eq:similarity}, it is known that $\mathbf{\Phi}_{2N}\mathbf{A}\mathbf{\Phi}^\dagger_{2N}=\mathbf{A}$ and $\mathbf{\Phi}_{2N}\mathbf{B}\mathbf{\Phi}^\dagger_{2N}=\mathbf{B}$. Hence, \eqref{eq:c_fresnel3} and \eqref{eq:c_fresnel4} become respectively
\begin{eqnarray}
\mathbf{\dot{c}} & = & \mathbf{A}\mathbf{\Phi}_{2N}\mathbf{b}\label{eq:c_fresnel5}\\
& = & \mathbf{B}\mathbf{\Phi}_{2N}\mathbf{a}\label{eq:c_fresnel6},
\end{eqnarray}
in which $\mathbf{\dot{a}}=\mathbf{\Phi}_{2N}\mathbf{a}$ and $\mathbf{\dot{b}}=\mathbf{\Phi}_{2N}\mathbf{b}$ are the Fresnel domain representations of $\mathbf{a}$ and $\mathbf{b}$, respectively.

From \eqref{eq:c_fresnel5} and \eqref{eq:c_fresnel6}, it is seen that the \ac{DFnT} of the circular convolution between $\mathbf{a}$ and $\mathbf{b}$ can be interpreted as the circular convolution between $\mathbf{a}$ and $\mathbf{\dot{b}}$ or, alternatively, as the circular convolution between $\mathbf{b}$ and $\mathbf{\dot{a}}$.

\bibliographystyle{IEEEtran}
\bibliography{referencias}

\end{document}